\documentclass[draft]{agujournal2019}
\usepackage{url} %this package should fix any errors with URLs in refs.
\usepackage{lineno}
\usepackage[inline]{trackchanges} %for better track changes. finalnew option will compile document with changes incorporated.
\usepackage{soul}
\usepackage{amsmath,amssymb}
\journalname{JGR: Space Physics}
\begin{document}

\title{Quantifying the Effects of Magnetic Field Line Curvature Scattering on Radiation Belt and Ring Current Particles}

\authors{Bin Cai\affil{1,2,3}, Hanlin Li\affil{1,2}, Yifan Wu\affil{1,2}, and Xin Tao\affil{1,2}}

\affiliation{1}{Deep Space Exploration Laboratory/School of Earth and Space Sciences, University of Science and Technology of China, Hefei, China}
\affiliation{2}{CAS Center for Excellence in Comparative Planetology/CAS Key Laboratory of Geospace Environment, University of Science and Technology of China, Hefei, China}
\affiliation{3}{Institute of Frontier and Interdisciplinary Science, Shandong University, Qingdao, China}

\correspondingauthor{X. Tao}{xtao@ustc.edu.cn}

%% Keypoints, final entry on title page.

%  List up to three key points (at least one is required)
%  Key Points summarize the main points and conclusions of the article
%  Each must be 140 characters or fewer with no special characters or punctuation and must be complete sentences

% Example:
% \begin{keypoints}
% \item	List up to three key points (at least one is required)
% \item	Key Points summarize the main points and conclusions of the article
% \item	Each must be 140 characters or fewer with no special characters or punctuation and must be complete sentences
% \end{keypoints}

\begin{keypoints}
\item The effects of field line curvature scattering on particles are investigated by test particle simulations and compared with theories. 
\item The decay time distributions of electrons and protons are calculated in T89 field model under different geomagnetic levels.
\item The effects of field line curvature scattering can be significant near $6\,R_\text{E}$ for MeV electrons and hundreds keV protons. 
\end{keypoints}

%% ------------------------------------------------------------------------ %%
%
%  ABSTRACT and PLAIN LANGUAGE SUMMARY
%
% A good Abstract will begin with a short description of the problem
% being addressed, briefly describe the new data or analyses, then
% briefly states the main conclusion(s) and how they are supported and
% uncertainties.

% The Plain Language Summary should be written for a broad audience,
% including journalists and the science-interested public, that will not have 
% a background in your field.
%
% A Plain Language Summary is required in GRL, JGR: Planets, JGR: Biogeosciences,
% JGR: Oceans, G-Cubed, Reviews of Geophysics, and JAMES.
% see http://sharingscience.agu.org/creating-plain-language-summary/)
%
%% ------------------------------------------------------------------------ %%

%% \begin{abstract} starts the second page

\begin{abstract}
Magnetic field line curvature (FLC) scattering is a collisionless scattering mechanism that arises when a particle's gyro-radius is comparable to the magnetic field line's curvature radius, resulting in the breaking of the conservation of the first adiabatic invariant. Studies in recent years have explored the implications of FLC scattering on the precipitation of both ring current ions and radiation belt electrons. In this work, we first compare two previous FLC scattering coefficients using test particle calculations. Then, we systematically calculate diffusion coefficients from FLC scattering in radial and MLT directions for particles of various energy levels, as well as its sensitivity to the $Kp$ index. We find that the timescale of FLC scattering is sufficient to account for the sudden loss of MeV electrons near the geostationary orbit during disturbed times. Additionally, the decay time of ring current protons is on the order of hours to minutes, providing an explanation for the ring current decay throughout the recovery phase of magnetic storms. Lastly, we compare the effects of wave-particle resonant scattering and FLC scattering in the vicinity of the midnight equator. Our findings suggest that the impacts of FLC scattering on MeV electrons or hundreds keV protons with smaller pitch angle is comparable to, or even more significant than, the effects of whistler mode or EMIC wave resonant scattering. Our quantitative results should be useful to evaluate the importance of the effects of FLC scattering while modeling the dynamics of radiation belt and ring current.
\end{abstract}

% \section*{Plain Language Summary}
% [ enter your Plain Language Summary here or delete this section]

\section{Introduction}
Charged particles in the inner magnetosphere have three periodic motions associated with three adiabatic invariants \cite{Northrop1964,Hastie1967,Roederer1970,Lyons1984}. We focus on the first adiabatic invariant in this work, which is $\mu = p^2 \sin^2\alpha/2m_0B$ . Here $\boldsymbol{p} $ is the momentum, $\alpha $ is the local pitch angle, $m_0$ is the mass, and $B$ is the local magnetic field strength. Conservation of $\mu $ requires that the ratio of the time scale of charged particle's cyclotron motion to field variation, as well as the ratio of the spatial scale of the charged particle gyro-radius $\rho$ to magnetic field inhomogeneity (or the radius of field line curvature $R_c$), be significantly less than one. Without electromagnetic waves, the ratio of spatial scale plays a critical role in determining whether the conservation of the magnetic moment is violated. Therefore, a parameter called the adiabatic parameter $\varepsilon = {\rho} / {R_c} $ has been used to determine the conservation of the first adiabatic invariant  \cite{Birmingham1982,Büchner1989,Anderson1997,Tu2014}. Field line curvature (FLC) scattering occurs when the radius of magnetic field line curvature is comparable to the gyro-radius of charged particles. 

Several analytical models have been established to predict the change of the first adiabatic invariant ($\Delta \mu$) due to FLC scattering \cite{Il'In1978,Birmingham1984,Delcourt1994,Anderson1997}. They found that for a single pass through minimum $B$, the change in $\mu$ has an exponential dependence on the adiabatic parameter $\varepsilon $, and is periodic in the gyro-phase $\phi_0$, where $\phi_0 $ is the particle gyro-phase angle at the $B$ minimum point. The jump of $\mu$ is in general of the form $\Delta \mu = A \cos\phi_0 \exp(-C/\varepsilon) $, $A $ and $C $ are coefficients related to magnetic field configuration and particle \cite{Young2002,Young2008}. Moreover, for the case where the trapped particle passes through the $B$ minimum multiple times, the cumulative effects of $\Delta\mu$ need to be taken into account. Given the condition that the individual random changes in $\mu $ is small and the cumulative effects are significant, the process of abrupt step-like changes of $\mu $ can be considered as the collisionless pitch angle scattering using the diffusion approximation \cite{Birmingham1984}. By driving the pitch angle distribution toward uniformity, the FLC scattering can cause the trapped particle into the loss cone to significantly affect the plasma transport in the absence of plasma waves. In the regime of terrestrial magnetosphere, the FLC scattering plays an important role in the location of the outer boundary of protons radiation belt \cite{Selesnick2023}, the rapid decay of ring current \cite{Ebihara2011,Chen2019,Yu2020}, the formation of proton isotropic boundary (IB) \cite{Gilson2012,Yue2014,Dubyagin2018,Ma2022}, and the regional enhancement of conductance of ionosphere \cite{Zhu2021}.

FLC scattering might also play a significant role in energetic electron scattering, despite electrons having a smaller gyro-radius than ions at the same kinetic energy. Based on a current sheet magnetic field model, \citeA{Artemyev2013} showed that the impact of FLC scattering is more significant than the scattering by wave-particle resonant interaction for relativistic electrons in the night side inner magnetosphere. According to \citeA{Eshetu2018}, global MHD simulations have shown that FLC scattering in the weak magnetic strength region adjacent to the BBFs can scatter energetic electrons with energies above a few keV. Using a multi-spacecraft analysis method, the study by \citeA{Zhang2016} offers in situ evidence of electron pitch angle scattering due to FLC scattering in the ion diffusion region of magnetic reconnection.

% \citeA{Eshetu2021} proposed an empirical model for the e-folding lifetime of midnight ions as a function of particle energy, pitch angle, L-value, and $Kp$ index in the T89c model to improve the inaccurate estimation of ring current ion loss. 

In this study, we first investigate and compare the accuracy of the \citeA{Birmingham1984} analytical model and \citeA{Young2002} empirical model for the calculation of FLC diffusion coefficients using test particle simulations. We then quantify the diffusion rates of FLC scattering of electrons and ions at different MLT, $L$-shell, and $Kp$-index values using the T89 magnetic field model. Additionally, we estimate the decay time of particles from FLC scattering and compare that from wave-particle resonant scattering. By doing so, we aim to provide a global picture about possible roles played by FLC scattering in the dynamics of radiation belt and ring current particles.

\section{Validation of FLC diffusion coefficients}
\label{sec:validation-analytic-model}

\subsection{Test particle simulation setup}
\label{sec:TP-simu}
% \subsection{Methodology}
Test particle simulations are used to investigate the detailed effects of FLC scattering on charged particles and to compare two sets of previously proposed FLC diffusion coefficients \cite{Birmingham1984,Young2002}. These simulations trace the complete trajectories of particles by solving the fully relativistic Lorentz motion equation using the Boris algorithm \cite{Birdsall2004}. The Boris algorithm is widely used due to its exceptional long-term accuracy \cite{Tao2011a,Fu2019,cai_effects_2020}. The fully relativistic Lorentz equations of motion are solved; i.e.,  
\begin{align}
& \frac{\mathrm{d} \boldsymbol{x}}{\mathrm{d} t}=\frac{\boldsymbol{p}}{\gamma m_0}, \\
& \frac{\mathrm{d} \boldsymbol{p}}{\mathrm{d} t}=q\left[\frac{\boldsymbol{p}}{\gamma m_0} \times\boldsymbol{B}\right],
\end{align}
where $\boldsymbol{x}$ is the particle position, and $\gamma$ is the Lorentz factor. Since we focus on FLC scattering, there is no electric field in the simulation. By analyzing the trajectories of individual particles and the ensemble behavior of particles over several bounce periods, quantitative diffusion coefficients from FLC scattering can be determined.

In the test particle simulation, we select a thin current sheet magnetic field as the background magnetic field model \cite{Wagner1979,Gray1982}, which is 
\begin{align}
  \label{eq:current-sheet-field}
   \boldsymbol{B} = B_{x0}\tanh(z/L_\text{CS}) \boldsymbol{e}_x + \sigma B_{x0} \boldsymbol{e}_z. 
\end{align}
Here $L_{\text{CS}}$ refers to the thickness of the current sheet, $B_{x0} $ represents the magnetic field strength at the boundary of the simulation where $z \gg L_{\text{CS}}$, and $\sigma $ is a constant parameter indicating the strength of magnetic field perturbation. The radius of curvature $R_c(z)$ can be easily obtained from equation (\ref{eq:current-sheet-field}). Specifically, at the center of the current sheet ($z=0$), we have $R_c(z=0) = \sigma L_{\text{CS}}$. The non-adiabatic behavior of particles with specific energy and equatorial pitch angle can be examined by varying the thicknesses of the current sheet $L_\text{CS} $. For our analysis, we assume $B_{x0}=100$\,nT and $\sigma=0.15$, following \citeA{Artemyev2013}. The time step used is $T_\text{g}/50$, where $T_\text{g} $ denotes the period of cyclotron motion of particles at the equator ($z=0$).

\begin{figure}
\centering
\noindent\includegraphics[width=0.8\textwidth]{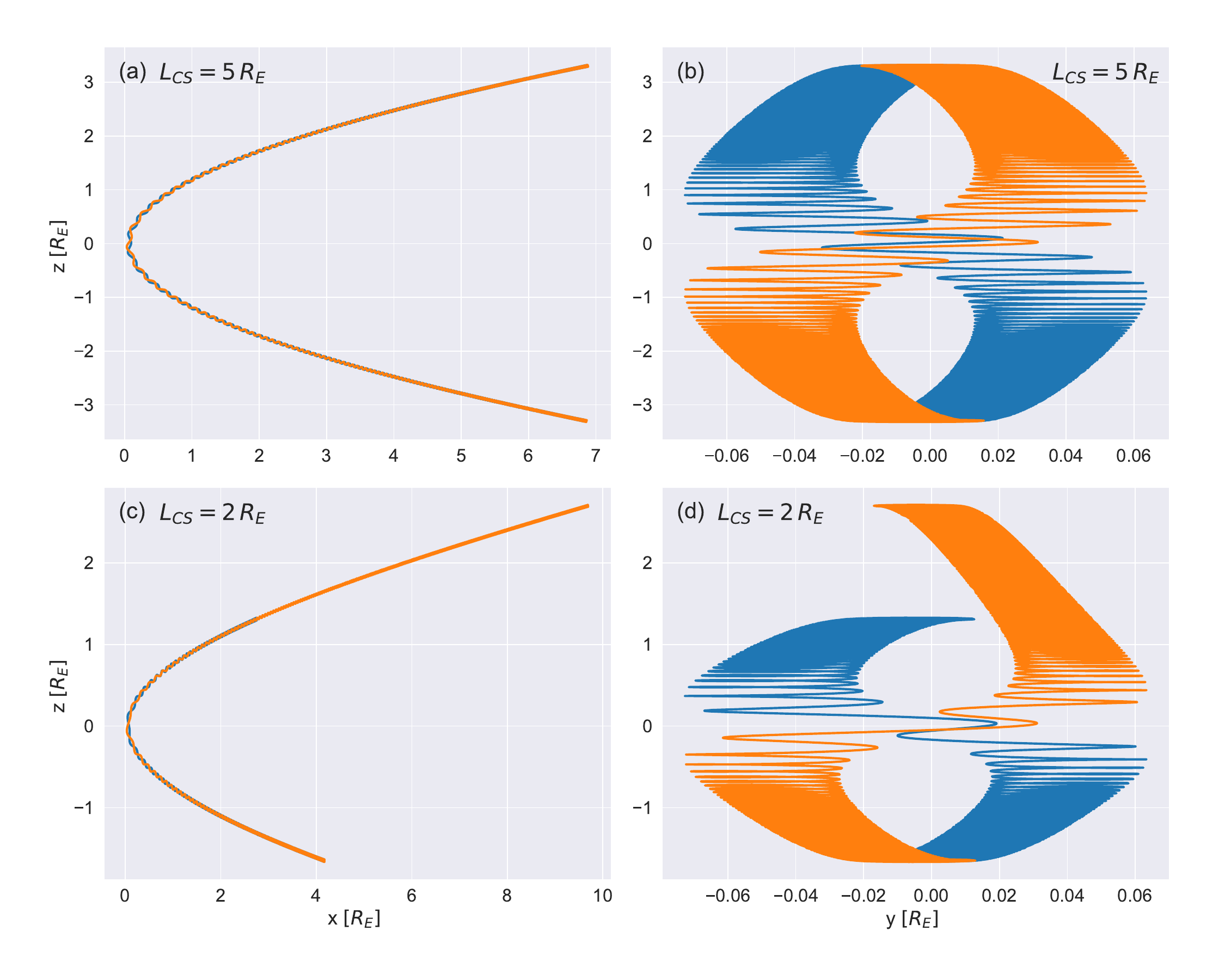}
\caption{Electron trajectories crossing the equator during a single bounce period are depicted in the one-dimensional current sheet field model. Figures (a) and (b) illustrate adiabatic motion corresponding to a current sheet thickness of $L_\text{CS} = 5 R_\text{E} $ and an adiabatic parameter of $ \varepsilon =0.09$. Figures (c) and (d) show non-adiabatic motion corresponding to a current sheet thickness of $L_\text{CS} = 2 R_\text{E} $ and an adiabatic parameter of $ \varepsilon =0.22$. Blue and orange indicate the trajectories of the electron from north to south and from south to north, respectively.}
\label{fig:1}
\end{figure}

Figure \ref{fig:1} shows electron trajectories from a single cross of minimum $B$ as a detailed example of FLC scattering. The initial energy $E$ of the electron is $1.5 $ MeV, and the initial equatorial pitch angle $\alpha_{0}$ is set to $ 30^\circ$. Two magnetic field configurations, $L_\text{CS} = 5\,R_\text{E}$ and $2\,R_\text{E}$, are used, and the corresponding adiabatic parameters are $\varepsilon = 0.09$ and $0.22$, respectively. Figure \ref{fig:1}(a) and \ref{fig:1}(c) represent the projections of trajectories in the $x$-$z$ plane, and Figure \ref{fig:1}(b) and \ref{fig:1}(d) represent the projections in the $y$-$z$ plane. The trajectories within one bounce period of electron initially moving from the northern magnetic mirror point to the southern is presented, illustrating paths in blue and orange for electron traveling from north to south and south to north, respectively, eventually returning to the northern magnetic mirror point. For $L_\text{CS} = 5\,R_\text{E}$, the electron's motion exhibits approximate field-aligned, and the trajectories display symmetry with respect to the $z=0$ plane. These results indicate the adiabatic behavior of electrons for $\varepsilon=0.09$. On the other hand, when $L_\text{CS} = 2\,R_\text{E}$, the electron trajectories show features typical of FLC scattering. The trajectories do not exhibit symmetry about the $z=0$ plane, and the equatorial pitch angle changes significantly, resulting in a considerable variation in the mirror point before and after the cross of the equatorial plane.

\subsection{Comparison of diffusion coefficients from multiple scattering}
\begin{figure}
\centering
\noindent\includegraphics[width=0.8\textwidth]{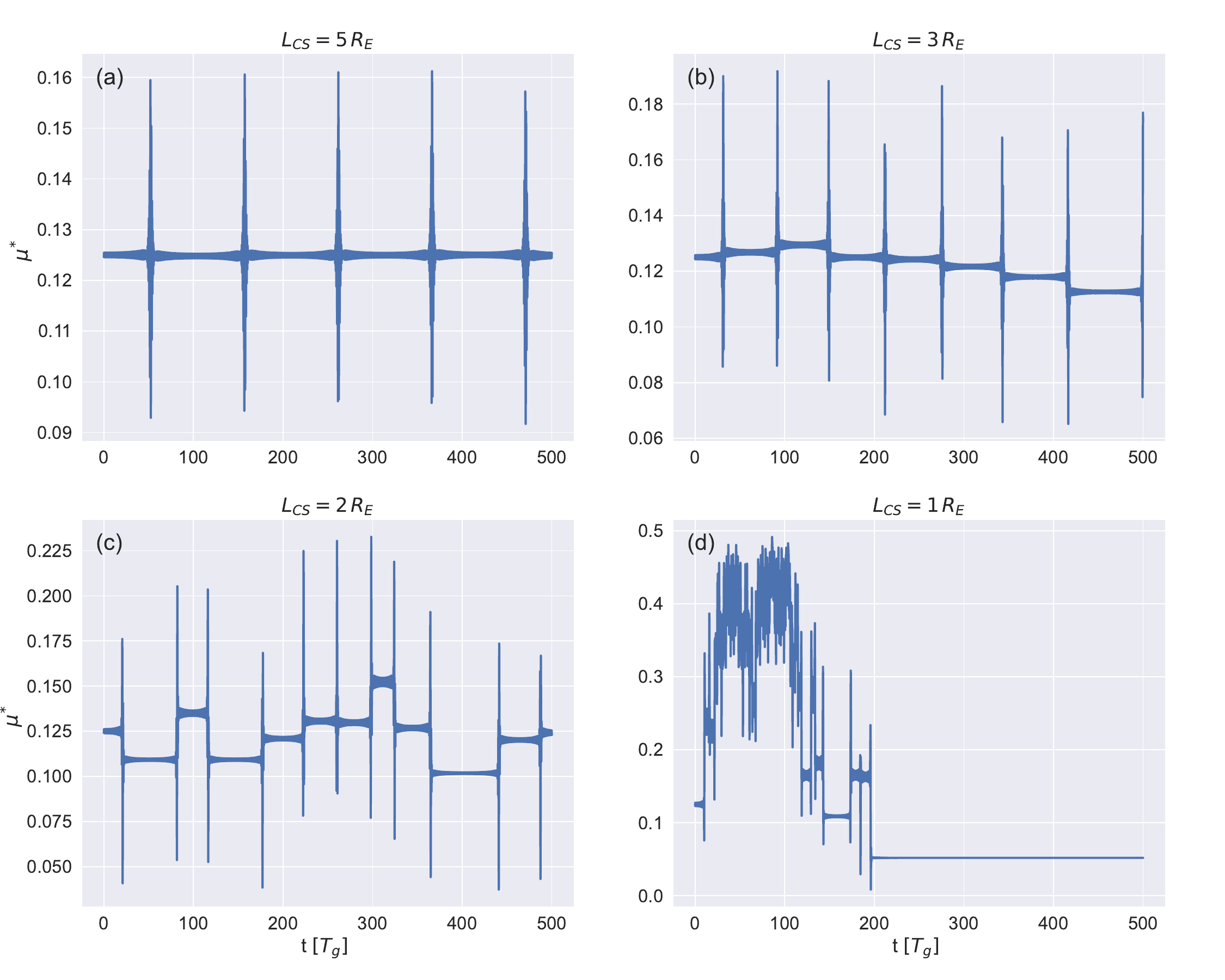}
\caption{The normalized magnetic moment exhibits time variation for four different current sheet field thicknesses, denoted as $ L_\text{CS}=5,\,3,\,2,\,1\,R_\text{E}  $, corresponding to $\varepsilon=0.09,\ 0.15,\ 0.22,\ 0.45 $, respectively.}
\label{fig:2}
\end{figure}

To compare the cumulative FLC scattering results obtained from test particle simulations with previous models, we increase the simulation time to 500 equatorial gyro-motion periods ($T_\text{g} $). The first adiabatic invariant, $\mu = p^{2}_{\perp}/(2m_0 B)$, is recorded at each time step for two additional magnetic field configurations: $L_\text{CS}= 3\,R_\text{E} $ and $1\,R_\text{E} $, corresponding to adiabatic parameters of $\varepsilon = 0.15 $ and $0.45 $, respectively. Here $p_{\perp} $ represents the particle momentum perpendicular to the magnetic field direction. Following previous studies \cite{Birmingham1984,Young2002}, we adopt the normalized first adiabatic invariant $\mu^{*} \equiv \mu / (p^2/m_{0} B) = \sin^2\alpha_{0} / {2} $ at the equatorial plane, with $\alpha_0 $ the equatorial pitch angle. For comparisons below with different field configurations, we use the same initial consistent equatorial pitch angle of $\alpha_{0} = 30^{\circ}$, and correspondingly, $\mu^{*} = 0.125$. The results are depicted in Figure \ref{fig:2}, illustrating the transition from adiabatic to non-adiabatic particle behavior. In the case of $L_\text{CS}=5\,R_\text{E} $, Figure \ref{fig:2}(a), the particle behaves adiabatically, as discussed above. The first adiabatic invariant remains nearly constant as the particle crosses the equatorial plane, although there is apparent rapid oscillation around the equatorial plane, which is due to that $\mu$ is not a strict invariant. Figures \ref{fig:2}(b) and \ref{fig:2}(c) display different degrees of non-adiabatic behavior caused by FLC scattering. A thinner current sheet thickness ($L_\text{CS}=2\,R_\text{E} $ corresponding to $\varepsilon=0.22 $) results in a more significant change in $\mu^{*}$, consistent with FLC scattering. Additionally, Figure \ref{fig:2}(d) shows a considerable change in $\mu^{*}$, several times larger than the initial value, which persists until the end of the simulation. The current sheet thickness in Figure \ref{fig:2}(d) is so thin ($L_\text{CS}=1\,R_\text{E} $) that the change in $\mu $ is significant when the electron crosses the equatorial plane and enters the loss cone after several bounce periods.

Although the FLC scattering effects induce non-adiabatic behavior in individual particles, its cumulative effects on variations in $\mu $ could be quantified through diffusion coefficients. Consequently, in addition to analyzing individual particle $\mu$ jump events, we investigate pitch angle scattering of energetic electrons and compute the diffusion rates of test particles. The diffusion coefficient ($ D^{\mathrm{TP}}_{\mu^* \mu^*}$) of the test particles' normalized first adiabatic invariant is determined using the expression $ D^{\mathrm{TP}}_{\mu^* \mu^*} \equiv \langle\Delta \mu^{*2}\rangle / 2 \Delta t $, where $ \langle \ldots \rangle $ denotes the average over all electrons, and $ \Delta \mu^* \equiv \mu^* - \langle \mu^* \rangle$ \cite{Tao2011a,Tao2012}. Subsequently, the equatorial pitch angle diffusion coefficient $D_{\alpha_0\alpha_0}$ can be derived using the relationship $D_{\alpha_0\alpha_0} = { D_{\mu^{*}\mu^{*}}}/({\sin^2\alpha_{0} \cos^{2}\alpha_{0}}) $. In order to calculate the diffusion coefficients of test particles, we track $200$ electrons for a given initial pitch angle $\alpha_0$ and energy. The initial gyro-phase is randomly assigned between 0 and 2$\pi $, and the electrons' initial $z $ position is uniformly distributed between two mirror points. We use the same initial energy of 1.5 MeV, but perform simulations for pitch angles between $30^{\circ}$ to $70^{\circ}$ with the step size of $1^{\circ}$.

\begin{figure}
\noindent\includegraphics[width=\textwidth]{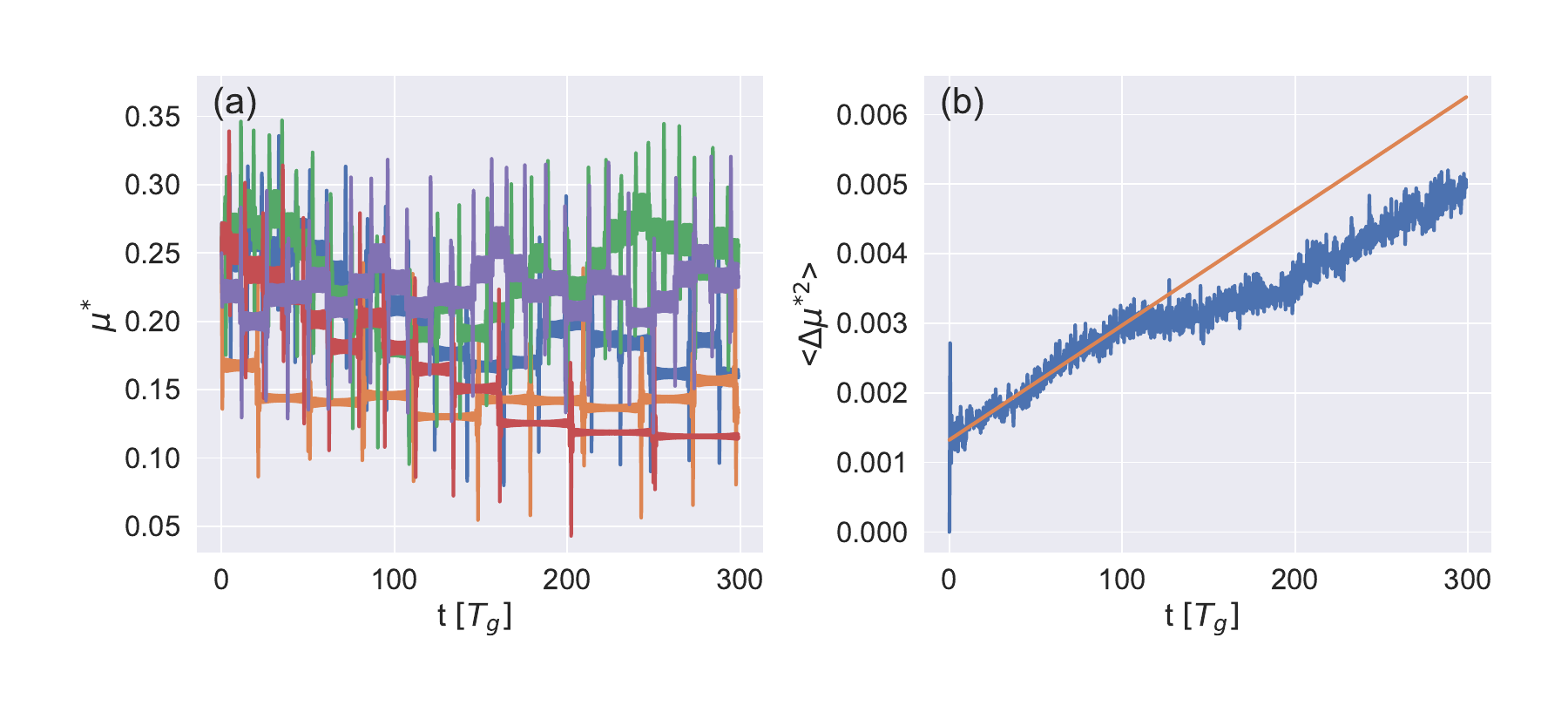}
\caption{(a) The normalized magnetic moment of five sampled electrons varies over time, all having the same initial equatorial pitch angle ($45^{\circ} $) and energy (1.5 MeV). (b) The orange line in the plot represents the linear fit of $\left\langle \Delta \mu^{*2} \right\rangle $ over the range of $100\,T_\text{g}$. }
\label{fig:3}
\end{figure}

Figure \ref{fig:3}(a) illustrates the $\mu^* $ variation of five randomly chosen electrons with an initial equatorial pitch angle of $\alpha_{0} = 45 ^{ \circ} $. The variation of $\mu^* $ caused by FLC scattering can be regarded as a stochastic process. Moreover, to compute the diffusion coefficients of test particles, we perform a linear fitting of $\left\langle \Delta \mu^{*2} \right\rangle $ against time, as demonstrated in Figure \ref{fig:3}(b). The $\left\langle \Delta \mu^{*2} \right\rangle $ gradually deviates from a linear function of $t$ after approximately 100 $T_\text{g} $ due to the gradual deviation of the pitch angle of scattered electrons from the initial value of $\alpha_{0} $. Hence, when conducting a linear fitting analysis of $\left\langle \Delta \mu^{*2} \right\rangle $, we limit the time range to be from $0$ to 100 $T_\text{g} $, as shown by the orange line in Figure \ref{fig:3}(b).

\begin{figure}
\centering
\noindent\includegraphics[width=0.6\textwidth]{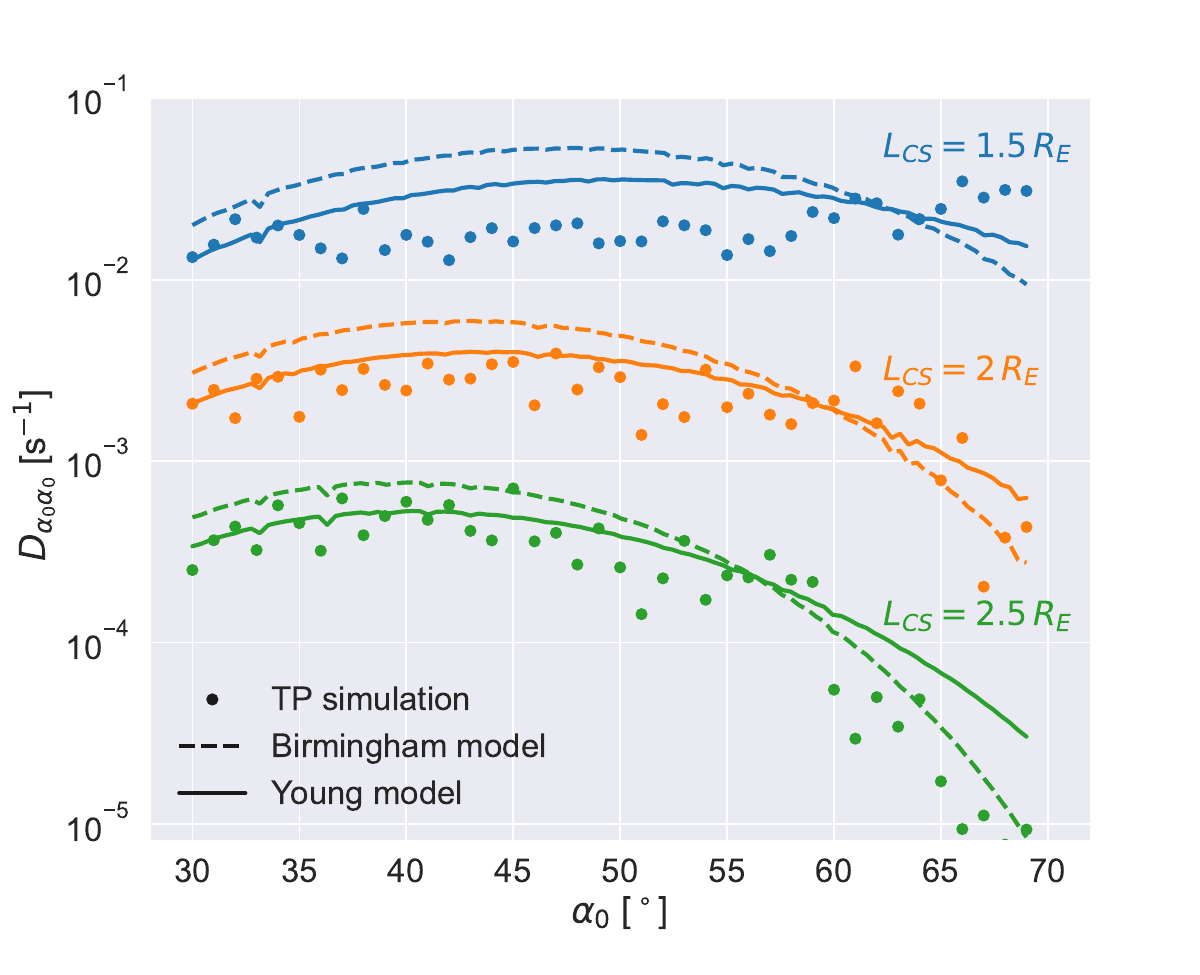}
\caption{The pitch angle diffusion coefficients as a function of equatorial pitch angle, ranging from $30^{\circ} $ to $70^{\circ}$, are computed from the simulations. The dashed and solid lines correspond to the results obtained from the \citeA{Birmingham1984} analytic model and \citeA{Young2002} empirical model, respectively. The three different current sheet thickness $L_\text{CS} $ conditions are distinguished by three different colors.}
\label{fig:4}
\end{figure}

Figure \ref{fig:4} presents a comparison between the pitch angle diffusion coefficients obtained from test particle simulations and diffusion coefficients from \citeA{Birmingham1984} and \citeA{Young2002} for three current sheet configurations, namely $L_\text{CS}= 1.5,\ 2,\ 2.5\,R_\text{E} $, with equatorial pitch angles ranging from $30^{\circ}$ to $70^{\circ}$. Overall, the \citeA{Birmingham1984} and \citeA{Young2002} models exhibits good agreement with the test particle simulations for $L_\text{CS}= 2 $ and $2.5\, R_\text{E} $ in the $30^{\circ}$ to $70^{\circ} $ pitch angle range. However, the \citeA{Young2002} model shows better consistency than \citeA{Birmingham1984} model at approximately $30^{\circ}$ to $60^{\circ}$. The lower diffusion coefficients from test particle simulations in the case of $L_\text{CS}= 1.5 R_\text{E} $ might be related to that this value of $L_\text{CS}$ corresponds to a thinner current sheet configuration than the other two cases. For equatorial pitch angles between $35^{\circ} $ and $ 60^{\circ} $, the change in $\mu^* $ of particle is significant over time, manifesting as the fast scattering of particles away from the initial $\mu^*$. Both factors lead to larger errors in test particle calculation of $\left\langle \Delta \mu^{*2} \right\rangle $. Overall, considering the better agreement with the test particle simulations, particularly within moderate equatorial pitch angle degrees, we select the \citeA{Young2002} empirical model as the tool to investigate the effects of FLC scattering in the inner magnetosphere below. 

\section{FLC scattering diffusion rate of energetic particles}
\label{sec:diffusion rate}
\subsection{Calculation of $D_{\alpha_0\alpha_0}$ using the T89 magnetic field model}
\label{sec-sub:calculate-Daa}
To investigate the impact of FLC scattering on relativistic electrons and energetic protons in the radiation belt and ring current, we employ the T89 magnetic field model as the background field \cite{Tsyganenko1989} with zero dipole tilt angle. In practice, the Python Geopack library, developed by \citeA{Tian}, is utilized for this. The T89 magnetic field model offers the advantages of being applicable under broad conditions and having simple parameter settings, making it well-suited for examining the global distribution of the diffusion rate resulting from FLC scattering.

Our purpose is to derive the 2D global distribution of diffusion coefficients on the $z=0$ plane in the T89 field using the \citeA{Young2002} empirical formula. These results will increase our overall understanding of the effects of FLC scattering on various particle kinds, such as electrons and protons, as a function of their energy, $L$-shell, MLT, and the geomagnetic activity index ($Kp$). The calculation process involves the following steps: (1) In the $z=0$ plane of GSM coordinates, we employ uniform 2D grids in the polar coordinate system, with 89 radial grids ranging from $1.2$ to $10\, R_\text{E}$ and 200 azimuthal grids ranging from $0$ to $2\pi$. Consequently, we calculate the diffusion coefficients at 17800 different locations, considering the particle energy and $Kp$ index. (2) Tracing the magnetic field line from the $z=0$ plane at each position to determine the magnetic field magnitude and the curvature radius at the B-minimum, which, in turn, are utilized to derive the adiabatic parameter $ \varepsilon $. (3) To determine the unperturbed bounce period $T_{b} $ of a particle with a specific equatorial pitch angle, we integrate the particle's parallel velocity along each field line. (4) The pitch angle diffusion coefficients are then calculated using the \citeA{Young2002} model. We select the initial equatorial pitch angle $\alpha_{0}=50^{\circ} $ as being representative due to the accurate estimation provided by the \citeA{Young2002} model for medium pitch angles, as demonstrated in Section \ref{sec:validation-analytic-model}.

\begin{figure}
  \centering
  \noindent\includegraphics[width=\textwidth]{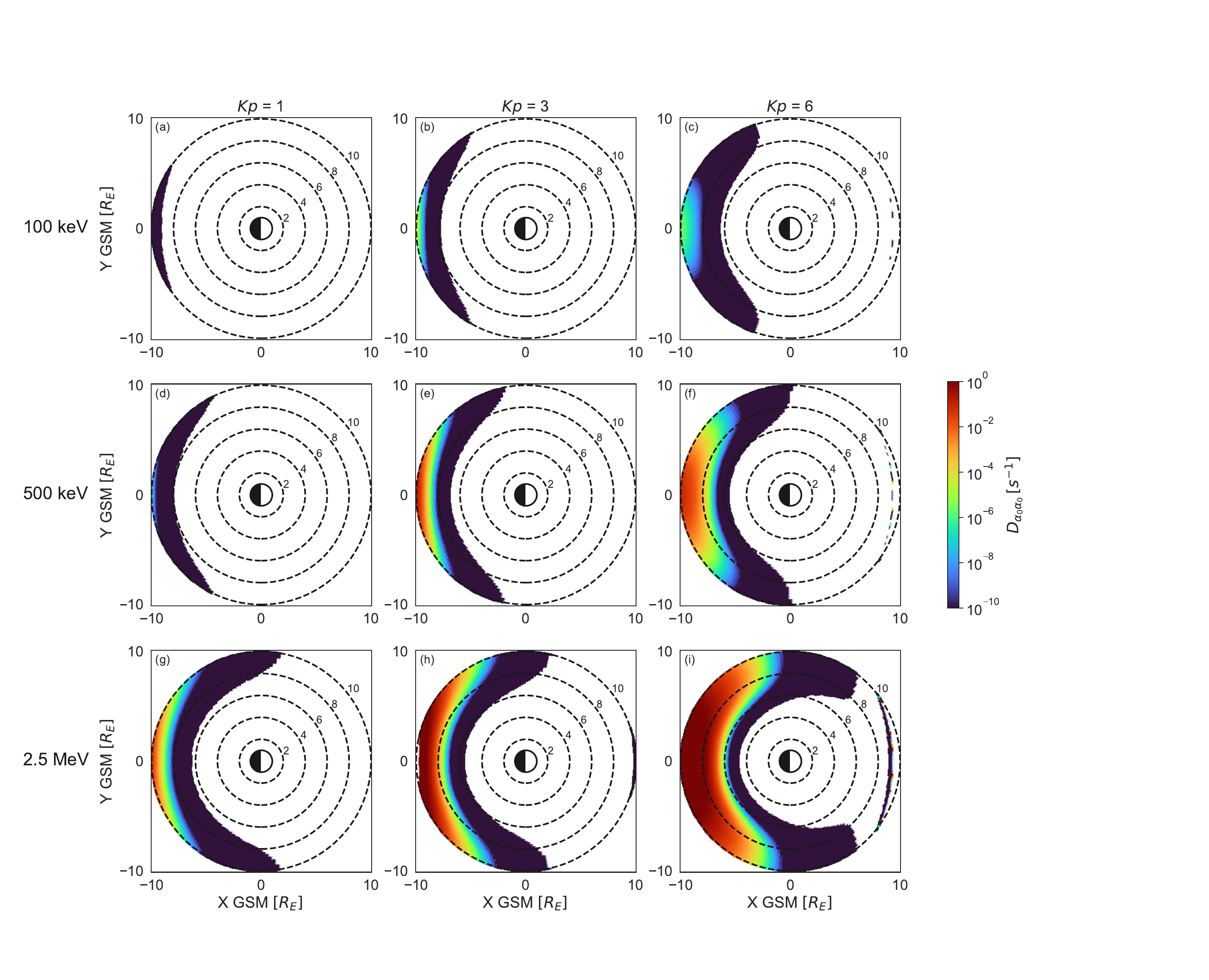}
  \caption{This figure illustrates the global distribution of diffusion coefficients induced by FLC scattering of electrons at the magnetic equatorial plane ($z=0$) in the T89 field. The rows correspond to three electron energy levels, energy equal to 100, 500 keV, and 2.5 MeV, while the columns represent three levels of geomagnetic activity, $Kp$ equals to 1, 3, and 6. The calculation extends up to an outer boundary of $10\,R_\text{E}$. }
  \label{fig:5}
\end{figure}

Figure \ref{fig:5} illustrates the pitch angle diffusion coefficients of electrons. Three energy levels, namely E = 100, 500 keV, and 2.5 MeV, are selected to represent outer radiation belt electrons. The figure consists of three columns, each representing different geomagnetic activity indices ($Kp$) equal to 1, 3, and 6, and three rows representing different energy levels. In general, as the electron energy increases and the geomagnetic activity becomes more intense, the FLC scattering region expands to encompass a broader range of $L$-MLT values. For radiation belt energetic electrons, the effects region of FLC scattering primarily occurs on the night side of the inner magnetosphere and rarely extends inward beyond the geostationary orbit ($L \sim 7 $), except for instances of high electron energy during extremely intense geomagnetic conditions. The observed regional distribution in our calculations is not unexpected. The stretched field lines primarily exist in the night-side inner magnetosphere, resulting in smaller curvature radii. Moreover, higher energy electrons possess larger gyro-radii. As a consequence, the adiabatic parameters tend to be greater than those obtained in the day-side magnetosphere or for lower electron energy levels. However, despite the rarity of FLC scattering occurrences within the geostationary orbit and their limitation to the night-side, the scattering coverage can extend to the dusk and lower portions of the inner magnetosphere, even crossing $L \sim 7 $ for electrons with an energy of 2.5 MeV and a $Kp$ index equals to 6. 

\begin{figure}
  \centering
  \noindent\includegraphics[width=\textwidth]{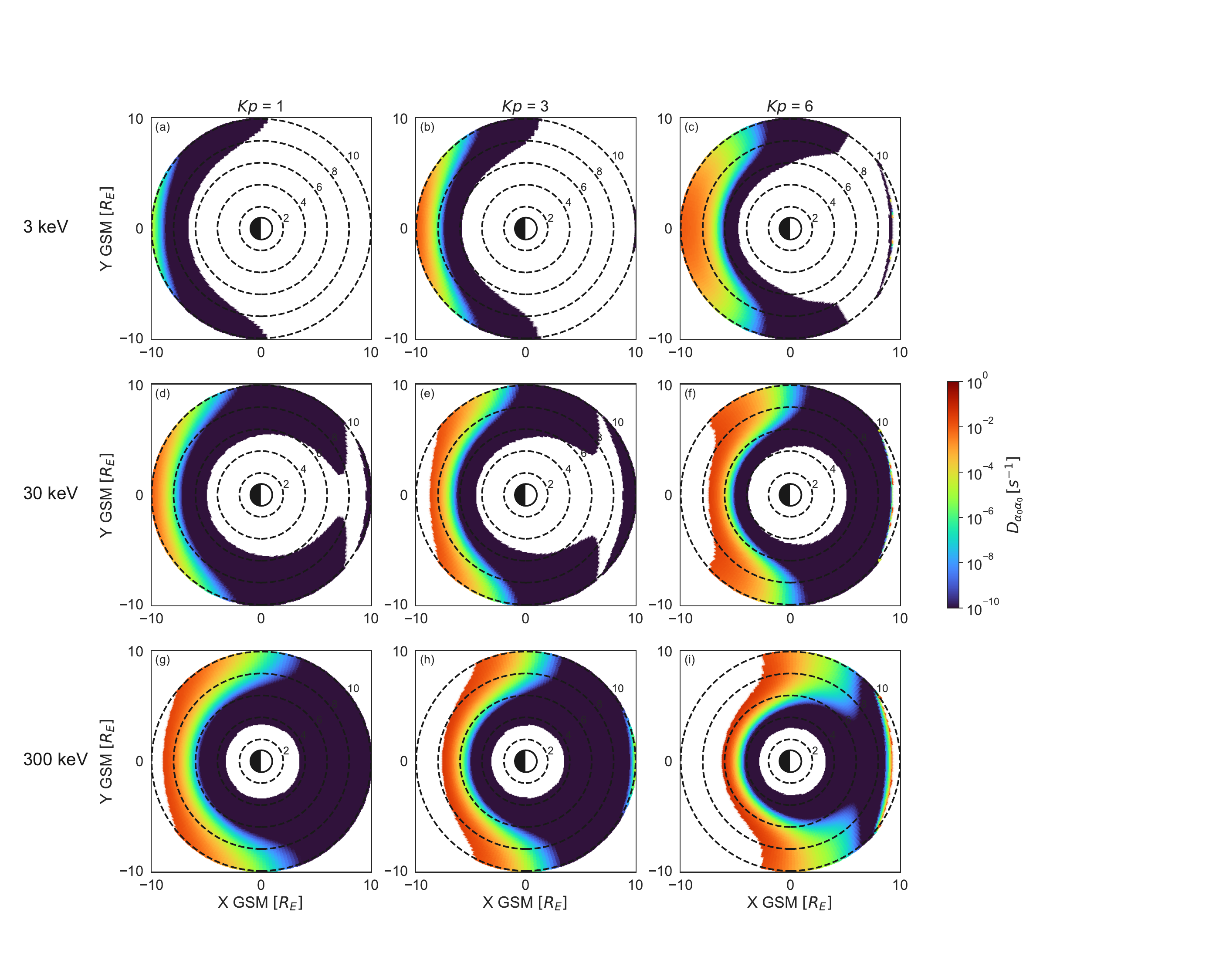}
  \caption{This study examines the global distribution of diffusion coefficients induced by FLC scattering for protons at the magnetic equatorial plane ($z=0$) in the T89 field. Proton energies of 3, 30, and 300 keV are considered.}
  \label{fig:6}
\end{figure}

Calculation of FLC diffusion coefficients is shown in Figure \ref{fig:6} for protons at three different energy levels: $E = 3$, 30 keV, and 300 keV, for different levels of geomagnetic activities. Based on the definition of the adiabatic parameter, the adiabatic parameter $\varepsilon $ is influenced not only by the magnetic field magnitude and curvature radius but also by the charge-to-mass ratio. Consequently, the diffusion of ions caused by FLC scattering is more likely to occur \cite{Tu2014,Yu2020,Eshetu2021,Zhu2021}. For energetic ions in the night side, the value of $\varepsilon $ becomes very large and even on the order of ten. We set the upper threshold of $\varepsilon $ to be 1, beyond which the multiple jump of $\mu$ is regarded as small chaotic scattering and the diffusion approximation can be invalid. As expected, for protons, the region of FLC scattering is considerably broader compared to electrons, even in the day-side magnetosphere, when the proton energy reaches hundred keV. For 3 keV and 30 keV protons, frequently observed in the ring current \cite{Sandhu2018,Chen2019}, the extent of FLC scattering is comparable to that of 2.5 MeV electrons in the MLT direction. The inner boundary of $D_{\alpha_0\alpha_0}>10^{-4} $ at midnight is located around $7-8 \,R_\text{E} $ for $Kp \le 4 $, and around 6 $ R_\text{E} $ for $Kp=6 $. In the case of 300 keV protons, the region affected by FLC scattering spans almost all MLT regions. The inner boundary extends inwardly to approximately $ 5\, R_\text{E} $, indicating that the protons will constantly experience FLC scattering as they drift around the Earth at a distance of $\sim 5 \,R_\text{E} $. 

\subsection{Decay time estimation}

Determining the time and spatial scale of a physical mechanism is crucial. In our study, it is necessary to determine the timescale for particle loss into the upper atmosphere caused by field line curvature scattering. Numerous studies have investigated the particle decay time in diffusion processes \cite{Kennel1966,Roberts1969,Schulz1974,Shprits2009,Albert2009}. In our study, we employ the approach proposed by \citeA{Shprits2006}, which has demonstrated that the decay time of electrons is highly influenced by the pitch angle scattering rate near the edge of the loss cone. It is important to note that the decay time represents the duration required for the total number of particles to decrease to $1/e$ of its initial value. Specifically, we define the height of particle precipitation loss as 1000 km above the surface and calculate the corresponding particle loss cone angle $\alpha_{L}$. Subsequently, we utilize the angle of the loss cone as the initial pitch angle of electrons at the magnetic equatorial plane to calculate the diffusion coefficients $D_{\alpha_0\alpha_0}^{*} \equiv D_{\alpha_0\alpha_0}(\alpha_0=\alpha_L)$ using the same method as described in section \ref{sec-sub:calculate-Daa}. Finally, by estimating the particle's decay time approximately using the value of $1/D_{\alpha_0\alpha_0}^{*}$ \cite{Shprits2006}, we obtain the timescale for the loss of electrons and protons.

\begin{figure}
  \centering
  \noindent\includegraphics[width=\textwidth]{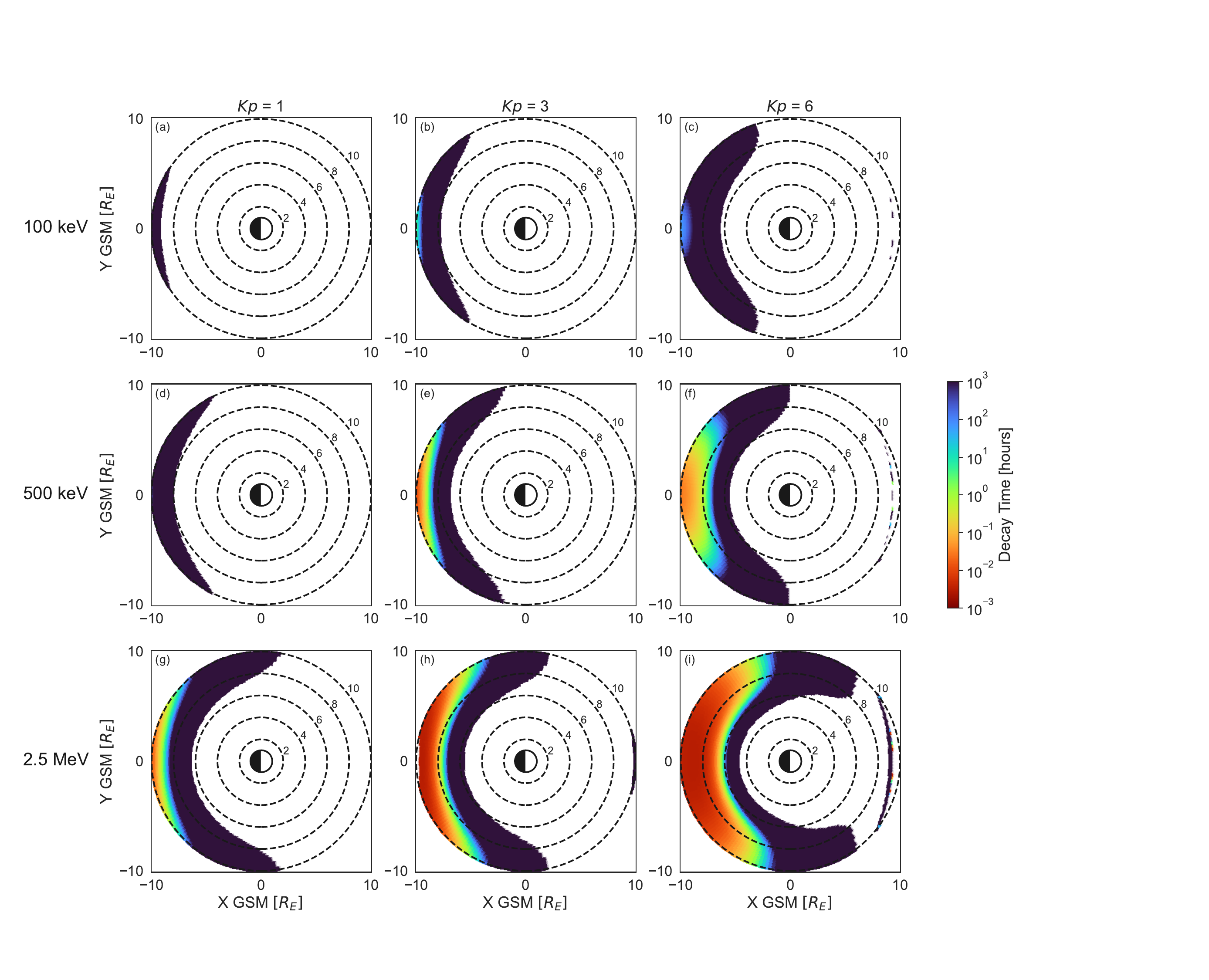}
  \caption{This figure illustrates the global distribution of the decay time (in hours) of electrons induced by FLC scattering in the T89 field. The remaining aspects are consistent with Figure \ref{fig:5}.}
  \label{fig:7}
\end{figure}

\begin{figure}
  \centering
  \noindent\includegraphics[width=\textwidth]{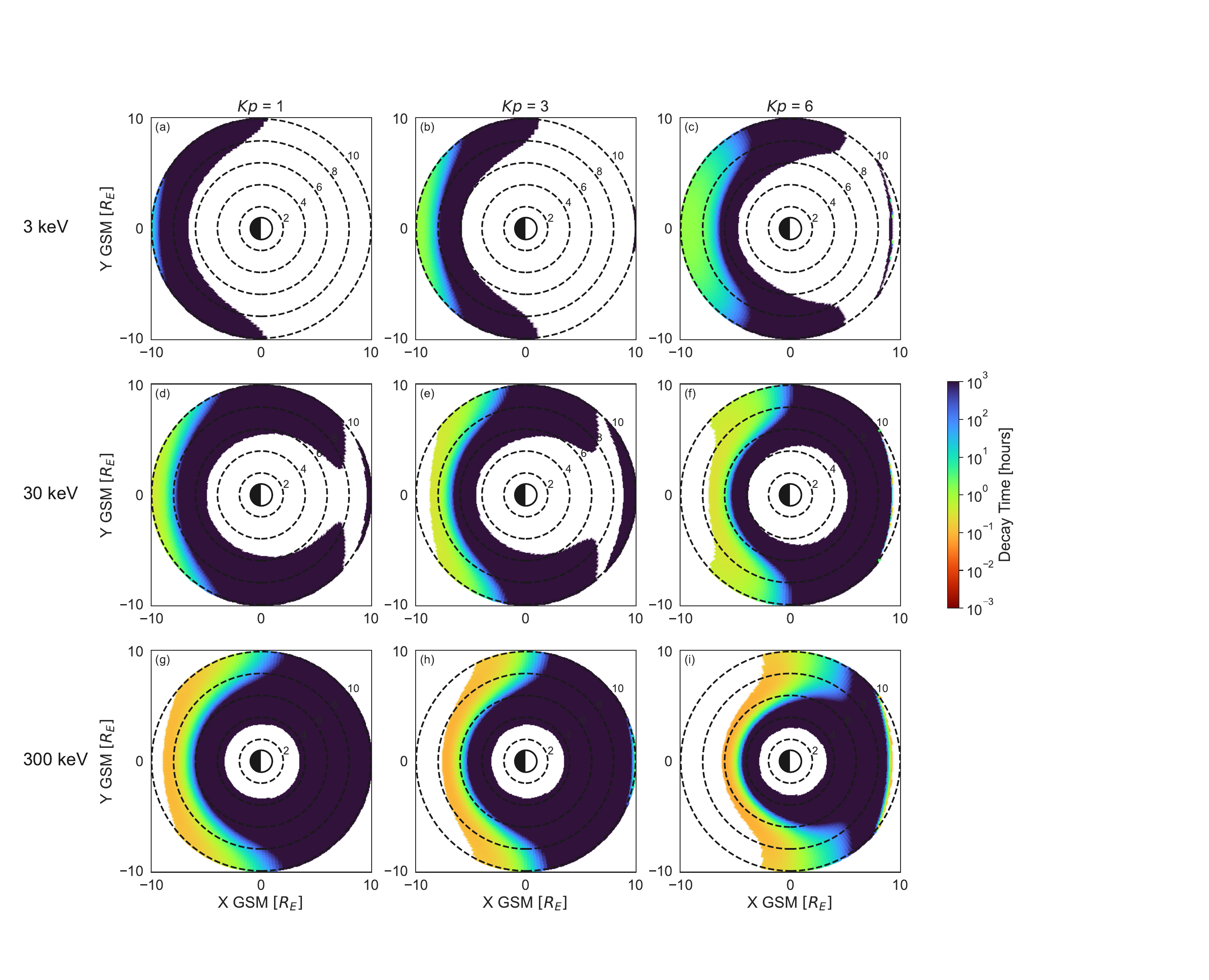}
  \caption{This figure presents the global distribution of the decay time (in hours) of protons induced by FLC scattering in the T89 field. The other aspects remain consistent with Figure \ref{fig:6}.}
  \label{fig:8}
\end{figure}

Figure \ref{fig:7} illustrates the decay time distribution of electrons, while Figure \ref{fig:8} depicts the decay time distribution of protons due to FLC scattering in the T89 model. Scattering processes with decay times exceeding 1000 hours (about 42 days) are considered insignificant and thus disregarded by leaving that region blank. The time of $10$, $10^{-1}$, $10^{-2}$ hours correspond to diffusion processes on the scale of days, minutes, and seconds, respectively. Regarding the decay time of electrons, the loss due to FLC scattering within $6\,R_\text{E}$ is ignorable for all three energies. However, for energy levels of 500 keV and 2.5 MeV, significantly increased loss occurs beyond $8\,R_\text{E}$ during intense geomagnetic activity with $Kp \ge 3$. Energy levels of 2.5 MeV electrons can even exhibit decay times on the scale of seconds just outside $6\,R_\text{E}$ under $Kp=6$ conditions. Overall, FLC scattering is unlikely to play a significant role in MeV electron dynamics at the heart of the outer radiation belt ($L\sim 4-5$). Regarding proton decay time, for energy levels of 30 keV, loss rates on the scale of hours occur at the night side between $6$ and $8\,R_\text{E}$ under $Kp=6$ conditions, which is comparable to the timescale of ring current decay during the storm's recovery phase. Protons with energy levels of 300 keV exhibit decay times of minutes around $6\,R_\text{E}$ under $Kp=6$ conditions, suggesting important roles of FLC scattering in proton dynamics during disturbed times. These calculations are also consistent simulation results reported by \citeA{Eshetu2021}. 

\subsection{Comparison with wave-particle resonant scattering}
\label{sec:compare-waves}
The wave-particle resonant interaction has long been a significant collision-less mechanism of interest for researchers studying the dynamic behavior of radiation belt electrons and ring current ions. For instance, whistler mode chorus waves can induce both local acceleration and the precipitation of electrons into the atmosphere through pitch angle and energy scattering processes \cite{Horne2003,Miyoshi2003,Ni2008,Thorne2010,Tao2011b}. Electromagnetic ion cyclotron (EMIC) waves not only result in the rapid loss of electrons with energies in the several MeV range but also induce ion precipitation with energies exceeding tens of keV \cite{Summers2003,Ebihara2011,Zhu2021}. It is necessary to compare the relative contributions of FLC scattering and wave-particle resonant interactions to particle loss. This comparison will increase our understanding of the charged particle dynamics in the inner magnetosphere. 

\begin{figure}
  \centering
  \noindent\includegraphics[width=\textwidth]{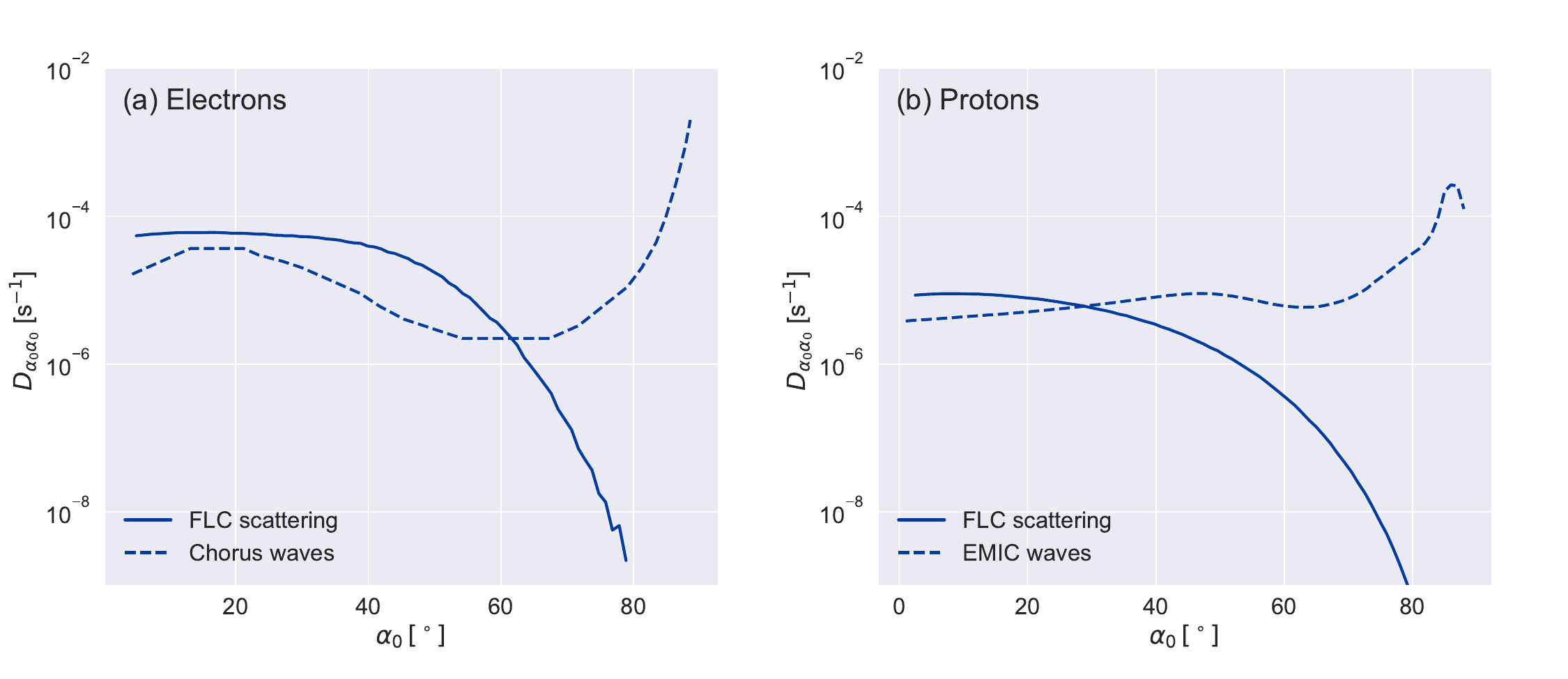}
  \caption{(a) The pitch angle diffusion coefficients $D_{\alpha_0\alpha_0}$ of relativistic electrons induced by FLC scattering (solid line) exhibit variations with equatorial pitch angle under the condition of $Kp$ = 6, at the midnight equatorial region of $7\,R_\text{E}$, following the same parameters as the work by \citeA{Orlova2010} (dashed line). (b) A comparison is made between the pitch angle diffusion coefficients from FLC scattering (solid line) and proton resonant scattering by electromagnetic ion cyclotron (EMIC) waves in the work of \citeA{Cao2016} (dashed line). The comparison is conducted at the midnight equatorial region of $8\,R_\text{E}$ under the T01 field, aligning with the parameters used in \citeA{Cao2016}. The energies of electrons and protons are 1 MeV and 100 keV, respectively.}
  \label{fig:9}
\end{figure}

\citeA{Orlova2010} computed the bounce-averaged diffusion coefficients of electrons induced by whistler mode chorus waves in the T89 field under different magnetic storm conditions. The whistler mode chorus waves have a Gaussian spread of wave power density, confined within magnetic latitudes $\pm$15$^\circ$, and wave amplitude $B_w =100$\,pT, following \citeA{Horne2005}. In Figure 1 of their paper (not shown here), the diffusion coefficients $D_{\alpha_{0}\alpha_0}$ for electron energies of 1\,MeV at the midnight location ($7\,R_\text{E}$) under a geomagnetic condition of $Kp=2$ and 6 are calculated. Here, we present our results of electron diffusion coefficients $D_{\alpha_0\alpha_0}$ induced by FLC scattering in Figure \ref{fig:9}(a), employing the method outlined in Section \ref{sec-sub:calculate-Daa}. We have chosen the same electron energy, location, and $Kp=6$ index for comparison. The $D_{\alpha_0\alpha_0}$ from FLC scattering is ignorable at $Kp=2$, and therefore, not compared. The comparison results show FLC scattering is more important for $\alpha_0 \lesssim 60^{\circ}$, while wave resonant interactions are more important for $\alpha_0 \gtrsim 60^{\circ}$. Therefore, FLC scattering could be the more important factor in particle loss at night side when the geomagnetic field is strongly perturbed ($Kp=6$). 

Another study we selected for comparison is the investigation of resonant interactions between parallel-propagating $\mathrm{H}^{+} $ band electromagnetic ion cyclotron (EMIC) waves and ring current protons conducted by \citeA{Cao2016}. They computed the diffusion coefficients for protons with energies of 10 keV and 100 keV using the T01 field \cite{T2002} at an $L$-value of $8\,R_\text{E} $. The frequency distribution, spectral intensity, and wave amplitude of EMIC waves are explained in detail by \citeA{Cao2016}. Here, we adopt the same input parameters of the T01 magnetic field model to compute the diffusion coefficients $D_{\alpha_0\alpha_0}$ induced by FLC scattering at the midnight location ($8\,R_\text{E}$). The results of the comparison are displayed in Figure \ref{fig:9}(b) for 100 keV protons. Again, we do not show the comparison for 10 keV protons because FLC scattering diffusion coefficients are ignorable for this energy. The overall conclusion is similar to that of electrons shown in Figure \ref{fig:9}(a); i.e., FLC scattering is more important at smaller pitch angles ($\alpha_0 \lesssim 30^{\circ}$) while wave-particle resonant interactions dominate at larger pitch angles ($\alpha_0 \gtrsim 30^{\circ}$). It is expected that, at higher proton energies, FLC scattering effects at the same location become more pronounced due to increased gyro-radius. Overall, FLC scattering for ring current protons should be important for the loss of high-energy ($>100$\,keV) protons at this location.
 
\section{Summary}
\label{sec:summary}
Charged particle scattering due to magnetic field line curvature (FLC) scattering plays a crucial role in non-adiabatic particle dynamics. Furthermore, it may contribute to the loss of energetic particles into the upper atmosphere. This study quantitatively investigates the effects of FLC scattering on radiation belt and ring current particles in a realistic magnetic field. We conducted fully relativistic test particle simulations to validate the accuracy of the analytical and empirical models proposed by \citeA{Birmingham1984} and \citeA{Young2002}, respectively. Using the empirical model by \citeA{Young2002}, we calculated the diffusion coefficients and decay time distribution of FLC scattering for both electrons and protons as a function of energy, local time, radial distance, and geomagnetic activity level. Additionally, we compare the relative significance of particle loss due to resonant scattering by waves and FLC scattering. The main findings can be summarized as follows:
\begin{enumerate}
  \item Electrons with energies about hundreds keV can experience the effects of FLC scattering at large distances $R > 8\,R_\text{E}$ and during the active period ($Kp = 6$) at midnight. In the case of ultra-relativistic electrons (several MeV), the FLC scattering region shifts closer to Earth $R = 6\,R_\text{E}$ and exhibits a wider area of importance in terms of electron loss on the night side, compared to sub-MeV electrons. Estimation of the decay time for 2.5 MeV electrons indicates that they can be lost within seconds to minutes.

  \item In comparison to electrons, protons are more significantly affected by FLC scattering due to a significantly larger gyro-radius. Protons with energies in the ring current range (hundreds of keV) can experience the effects of FLC scattering at distances less than $6\,R_\text{E}$ during the active period $Kp = 6$. The loss of these protons can occur within minutes, which is comparable to the timescale of ring current decay during the recovery phase of a magnetic storm. Hence, the FLC scattering can serve as an efficient mechanism for the precipitation of ring current protons during disturbed periods.

  \item In the region near $R = 7\,R_\text{E}$ and at $Kp=6$, the relative contribution of FLC scattering to MeV electrons is comparable to that of chorus waves at small pitch angles, and therefore, may contribute to electron loss. In contrast, at high pitch angles, FLC scattering is ignorable compared with wave-particle interactions for the parameters used in the comparison. Similar conclusions have been reached for ring current protons with energies of 100 keV at midnight $R = 6\,R_\text{E}$ under $Kp = 6$.
\end{enumerate}

Our findings should provide a global picture regarding the effects of FLC scattering in the particle dynamics near or within the inner magnetosphere; therefore, it should be valuable to determine when or where we should include FLC scattering in studies about the loss or scattering of radiation belt electrons and ring current protons.

\section*{Data Availability Statement}
The simulation data produced by this work can be found at https://doi.org/10.5281/zenodo.10215251. 

\acknowledgments
This work was supported by the National Key R\&D Program of China (2022YFF0503702) and NSFC grant 42174182. 

\bibliography{ustc}

%Reference citation instructions and examples:
%
% Please use ONLY \cite and \citeA for reference citations.
% \cite for parenthetical references
% ...as shown in recent studies (Simpson et al., 2019)
% \citeA for in-text citations
% ...Simpson et al. (2019) have shown...
%
%
%...as shown by \citeA{jskilby}.
%...as shown by \citeA{lewin76}, \citeA{carson86}, \citeA{bartoldy02}, and \citeA{rinaldi03}.
%...has been shown \cite{jskilbye}.
%...has been shown \cite{lewin76,carson86,bartoldy02,rinaldi03}.
%... \cite <i.e.>[]{lewin76,carson86,bartoldy02,rinaldi03}.
%...has been shown by \cite <e.g.,>[and others]{lewin76}.
%
% apacite uses < > for prenotes and [ ] for postnotes
% DO NOT use other cite commands (e.g., \citet, \citep, \citeyear, \nocite, \citealp, etc.).
%

\end{document}